\documentclass[journal,draftclsnofoot,onecolumn]{IEEEtran}

\usepackage{graphicx,amsfonts,amsthm}
\usepackage[cmex10]{amsmath}
\usepackage{verbatim,cite}

\def\tP{{\tilde{P}}}
\def\tR{{\tilde{R}}}
\def\tpi{{\tilde{\pi}}}
\def\mbpi{{\boldsymbol{\pi}}}
\def\mba{{\mathbf{a}}}
\def\mbc{{\mathbf{c}}}
\def\hH{{\hat{H}}}
\def\Bayes{{\mathrm{B}}}
\DeclareMathOperator*{\argmin}{\arg\,\min}
\DeclareMathOperator*{\cone}{cone}
\newtheorem{theorem}{Theorem}
\newtheorem{lemma}{Lemma}
\newtheorem{cor}{Corollary}
\theoremstyle{remark}
\newtheorem*{remark}{Remark}

\title{Robust Binary Hypothesis Testing Under Contaminated Likelihoods}
\author{Dennis Wei and Kush R.\ Varshney} 
\begin{document}
%
\maketitle
\begin{abstract}
In hypothesis testing, the phenomenon of label noise, in which hypothesis labels are switched at random, contaminates the likelihood functions.  In this paper, we develop a new method to determine the decision rule when we do not have knowledge of the uncontaminated likelihoods and contamination probabilities, but only have knowledge of the contaminated likelihoods.  In particular we pose a minimax optimization problem that finds a decision rule robust against this lack of knowledge.  The method simplifies by application of linear programming theory.  Motivation for this investigation is provided by problems encountered in workforce analytics.
\end{abstract}
\begin{keywords}
label noise, linear programming, minimax, signal detection theory, workforce analytics
\end{keywords}

\section{Introduction}
\label{sec:intro}

Label noise in hypothesis testing problems results in the cross-contamination of the likelihood functions and possible degradation in detection performance if not accounted for when determining a decision rule.  In this paper, we propose a linear programming framework for robustly dealing with contaminated likelihoods.  Specifically, we propose an algorithm for obtaining 
a minimax optimal decision rule under label noise that is applicable under general likelihood models.

We are motivated by problems encountered in workforce analytics: data-driven decision making to manage the human capital of a corporation.  For example, decision makers may want to use human resources data to predict whether or not an employee will voluntarily resign within the next 12 months \cite{SinghVWMGFE2012}, or decision makers may want to determine whether an employee from another division is a suitable candidate to fill an open position on a team in their division, based on skills and expertise data about the employee.  We face label noise and contamination of hypotheses in both examples.  In the voluntary resignation example, we can take all employees that resigned in the recent past as samples from the alternative hypothesis and all employees that are currently active as samples from the null hypothesis.  However, among currently active employees, some will resign in the coming months.  Therefore, we are not in a position to observe an uncontaminated null distribution.  In the suitable candidate example, we can take all employees in the decision maker's team as samples from the alternate distribution and all other employees as samples from the null distribution.  However, not all team members may be suitable for the open position and not all other employees are unsuitable (which is why this problem is posed in the first place).  Thus in this example, we observe contaminated versions of both likelihoods.

The problem of contaminated likelihoods in binary hypothesis testing was recently studied in considerable generality in \cite{scott2013COLT,scott2013arXiv}.  The theoretical framework in the present work is largely guided by \cite{scott2013COLT,scott2013arXiv}.  These previous works assume that the true likelihoods have an irreducibility property (described more fully in Section~\ref{sec:theory}) that allows consistency results to be established.  However, the assumption of irreducibility is restrictive.  It is not satisfied for example by two Gaussian distributions with different variances, nor is it likely to be satisfied by real-world distributions such as may be encountered in workforce analytics.  A contribution of the current paper in Section~\ref{sec:theory} is to remove the irreducibility assumption and extend the analysis to arbitrary true likelihoods.  Furthermore, the approach taken herein, described in Section~\ref{sec:robust}, differs fundamentally from \cite{scott2013COLT,scott2013arXiv} in focusing not on consistent learning of a particular contamination model, but rather on designing hypothesis tests that are robust to uncertainty in the model.  In Section~\ref{sec:numerical}, the utility of the robust viewpoint is demonstrated in two numerical examples.

More broadly, various types of label noise have been studied in the machine learning literature, including random, adversarial, and observation-dependent, and noise that affects different classes symmetrically and asymmetrically \cite{FrenayK2015}.  However, the vast majority of that work has been devoted to classifiers learned from finite training data and has been specific to particular 
supervised classification algorithms, see numerous references given in \cite{scott2013COLT,scott2013arXiv}.  In contrast, our work deals with the regime encountered in signal detection theory and hypothesis testing, not the regime with finite training samples.  Therefore, we work with likelihood ratio tests and true error probabilities rather than with specific classification algorithms and generalization bounds.  Somewhat more related is the mixture modeling approach of \cite{lawrence2001,bouveyron2009}, which attempts to learn the contamination model using the EM algorithm.  This approach however requires parametric assumptions on the true likelihoods that we do not make.

\section{Problem Statement}
\label{sec:prob}

We consider the binary hypothesis testing problem of deciding between a null hypothesis $H = h_0$ and an alternative hypothesis $H = h_1$ based on observation of a random variable $Y$.  Under hypothesis $H = h_0$, $Y$ follows the probability distribution $P_0$, while under $H = h_1$, $Y$ follows distribution $P_1$.  A decision rule $\hH$ is desired that maps every possible observation $Y = y$ to either $h_0$ or $h_1$.  For a rule $\hH$, define $R_0(\hH) = \Pr(\hH = h_1 \mid H = h_0)$ and $R_1(\hH) = \Pr(\hH = h_0 \mid H = h_1)$ to be the Type I and Type II error probabilities.  In this paper we focus on the Bayesian formulation in which the hypotheses have prior probabilities $\Pr(H = h_0) = q_0$, $\Pr(H = h_1) = 1-q_0$, and the performance measure is the Bayes risk 
\begin{equation}\label{eqn:BayesRisk}
R_B(\hH) = c_{01} q_0 R_0(\hH) + c_{10} (1-q_0) R_1(\hH),
\end{equation}
where $c_{01}$ and $c_{10}$ are the costs of Type I and Type II errors.

Given knowledge of the conditional distributions $P_0$ and $P_1$, it is straightforward to construct a likelihood ratio test that minimizes the Bayes risk \cite{VanTrees1968}.  However, in the contaminated version of the problem considered herein, $P_0$ and $P_1$ are not known. 
Instead, we have access to the contaminated distributions 
\begin{subequations}\label{eqn:Ptilde01}
\begin{align}
\tP_0 &= (1 - \pi_0) P_0 + \pi_0 P_1,\label{eqn:Ptilde0}\\
\tP_1 &= (1-\pi_1) P_1 + \pi_1 P_0,\label{eqn:Ptilde1}
\end{align}
\end{subequations}
where the contamination proportions $\pi_0, \pi_1 \in [0,1]$ are also unknown.  The following constraint is placed on $\pi_0$, $\pi_1$, 
\begin{equation}\label{eqn:contamBound}
\pi_0 + \pi_1 < 1,
\end{equation}
to resolve an interchange ambiguity and with essentially no loss of generality. 
Indeed, if $\pi_0 + \pi_1 > 1$, then as noted in \cite{scott2013COLT}, interchanging $P_0$ and $P_1$ yields complementary proportions $1-\pi_0$, $1-\pi_1$ satisfying $(1-\pi_0) + (1-\pi_1) < 1$.  If $\pi_0 + \pi_1 = 1$, then \eqref{eqn:Ptilde01} implies that $\tP_0 = \tP_1$ and discrimination is not possible. 

As discussed in \cite{scott2013COLT}, it is not possible in general to design a test $\hH$ that minimizes the Bayes risk \eqref{eqn:BayesRisk}, defined in terms of the true distributions $P_0$, $P_1$, given only the contaminated distributions $\tP_0$, $\tP_1$ and no knowledge of $P_0$, $P_1$, $\pi_0$, $\pi_1$.  Therefore in this paper we revise the objective to that of choosing $\hH$ 
to be robust to the uncertainty in 
$P_0$, $P_1$, subject to limited additional input.  
We note that in the absence of further conditions, 
there is a large range of possible solutions to \eqref{eqn:Ptilde01}.  In particular, it cannot be ruled out that there is no contamination, i.e.~$\pi_0 = \pi_1 = 0$, $P_0 = \tP_0$, and $P_1 = \tP_1$.  In the sequel, we seek to identify conditions that require minimal knowledge of or assumptions on $P_0$, $P_1$, $\pi_0$, $\pi_1$ while also restricting uncertainty in a meaningful way in terms of Bayes risk.

We focus in this paper on the population setting where the distributions $\tP_0$ and $\tP_1$ are known exactly.  Our results can be extended fairly straightforwardly to the finite-sample setting where $\tP_0$ and $\tP_1$ are approximated using training data, for example following the learning-theoretic approach of \cite{scott2013COLT}. 
In the finite-sample case, the lack of knowledge of $P_0$, $P_1$ translates into an inability to draw samples from $P_0$, $P_1$.

\section{Contamination Model Theory}
\label{sec:theory}

In this section we present 
results that precisely characterize the possible solutions $(P_0, P_1, \pi_0, \pi_1)$ to the contamination model \eqref{eqn:Ptilde01}.  These results generalize parallels in \cite{scott2013COLT} as discussed shortly.  

First we recall some definitions from \cite{scott2013COLT}.  For probability distributions $P$ and $Q$, define the maximal mixture proportion $\nu^\ast(P,Q)$ as 
\begin{equation}\label{eqn:maxMixProp}
\nu^\ast(P,Q) = \max \{\alpha \in [0,1] : \exists \text{ probability distribution } S : P = \alpha Q + (1-\alpha)S\}.
\end{equation}
One way of interpreting $\nu^\ast(P,Q)$ is as the infimum of the ratio $p(x)/q(x)$ if $P$ and $Q$ have probability densities $p(x)$ and $q(x)$ \cite[Lem.~5]{scott2013arXiv}.  From this it can be seen that $\nu^\ast(P,Q)$ is not necessarily symmetric.  If $\nu^\ast(P,Q) = 0$, $P$ is said to be irreducible with respect to $Q$, and if $\nu^\ast(Q,P) = 0$ also, then $P$ and $Q$ are mutually irreducible.  Many of the results in \cite{scott2013COLT} depend on the assumption that the true distributions $P_0$ and $P_1$ are mutually irreducible.  This assumption is relaxed in the present paper.

The first result below relates maximal mixture proportions between $P_0$ and $P_1$ to mixed counterparts involving both pure and contaminated distributions.
\begin{lemma}\label{lem:maxMixProp}
Under condition \eqref{eqn:contamBound},
\begin{align*}
\nu^\ast(P_0, \tP_1) &= \frac{\nu^\ast(P_0, P_1)}{1 - \pi_1 + \pi_1 \nu^\ast(P_0, P_1)},\\
\nu^\ast(P_1, \tP_0) &= \frac{\nu^\ast(P_1, P_0)}{1 - \pi_0 + \pi_0 \nu^\ast(P_1, P_0)}.
\end{align*}
\end{lemma}
\begin{IEEEproof}
It is shown that a decomposition of $P_0$ in terms of $P_1$ and another distribution $Q$ implies a decomposition of $P_0$ in terms of $\tP_1$ and $Q$, and vice versa.  Combining the implications yields the first equality in the lemma. The proof of the second equality is entirely analogous.

For the forward implication, let $\nu$ and $Q$ be such that 
\begin{equation}\label{eqn:maxMixProp1}
P_0 = \nu P_1 + (1-\nu) Q,
\end{equation}
where $\nu \leq \nu^\ast(P_0, P_1)$ by definition \eqref{eqn:maxMixProp}.  Given \eqref{eqn:contamBound}, \eqref{eqn:Ptilde1} can be solved for $P_1$ and the result substituted into \eqref{eqn:maxMixProp1} to yield 
\begin{align}
P_0 &= \nu \left( \frac{1}{1-\pi_1} \tP_1 - \frac{\pi_1}{1-\pi_1} P_0 \right) + (1-\nu) Q,\nonumber\\
P_0 &= \frac{\nu}{1-\pi_1+\nu\pi_1} \tP_1 + \frac{(1-\pi_1)(1-\nu)}{1-\pi_1+\nu\pi_1} Q.\label{eqn:maxMixProp2}
\end{align}
Since the numerators in \eqref{eqn:maxMixProp2} are non-negative and their sum equals the denominator, \eqref{eqn:maxMixProp2} is a valid mixture decomposition of $P_0$ in terms of $\tP_1$ and $Q$. It follows from \eqref{eqn:maxMixProp} that 
\begin{equation}\label{eqn:maxMixProp3}
\nu^\ast(P_0, \tP_1) \geq \frac{\nu}{1-\pi_1+\nu\pi_1}.
\end{equation}
Using the formula 
\begin{equation}\label{eqn:linFracDeriv}
\frac{d}{dx} \frac{Ax + B}{Cx + D} = \frac{AD-BC}{(Cx + D)^2},
\end{equation}
it is seen that the right-hand side of \eqref{eqn:maxMixProp3} is increasing in $\nu$.  Therefore the bound \eqref{eqn:maxMixProp3} is optimized at $\nu = \nu^\ast(P_0, P_1)$:
\begin{equation}\label{eqn:maxMixProp4}
\nu^\ast(P_0, \tP_1) \geq \frac{\nu^\ast(P_0, P_1)}{1 - \pi_1 + \pi_1 \nu^\ast(P_0, P_1)}.
\end{equation}

For the reverse implication, suppose that $P_0 = \nu \tP_1 + (1-\nu) Q$ for $\nu \leq \nu^\ast(P_0, \tP_1)$ and some $Q$.  Substituting for $\tP_1$ using \eqref{eqn:Ptilde1} and re-solving for $P_0$ as above gives
\begin{equation}\label{eqn:maxMixProp5}
P_0 = \frac{\nu(1-\pi_1)}{1-\nu\pi_1} P_1 + \frac{1-\nu}{1-\nu\pi_1} Q,
\end{equation}
which is again a valid mixture decomposition with non-negative coefficients that sum to $1$.  Furthermore, the coefficient in front of $P_1$ is increasing in $\nu$.  The combination of \eqref{eqn:maxMixProp} and \eqref{eqn:maxMixProp5} with the maximizing choice $\nu = \nu^\ast(P_0, \tP_1)$ implies 
\[
\nu^\ast(P_0, P_1) \geq \frac{(1-\pi_1) \nu^\ast(P_0, \tP_1)}{1-\pi_1 \nu^\ast(P_0, \tP_1)}.
\]
Solving the last inequality for $\nu^\ast(P_0, \tP_1)$ yields \eqref{eqn:maxMixProp4} but with the inequality reversed, completing the proof.
\end{IEEEproof}
Lemma~\ref{lem:maxMixProp} generalizes \cite[Lem.~3]{scott2013COLT}, which states that $\nu^\ast(P_0, \tP_1) = 0$ 
if and only if $\nu^\ast(P_0, P_1) = 0$, 
and similarly for the second equation.  
In the non-irreducible case, it can be seen that the maximal mixture proportion must increase with contamination according to the bounds below.
\begin{cor}
Under condition \eqref{eqn:contamBound}, 
\begin{align*}
\nu^\ast(P_0, P_1) &\leq \nu^\ast(P_0, \tP_1) \leq \frac{\nu^\ast(P_0, P_1)}{1 - \pi_1},\\
\nu^\ast(P_1, P_0) &\leq \nu^\ast(P_1, \tP_0) \leq \frac{\nu^\ast(P_1, P_0)}{1 - \pi_0}.
\end{align*}
Equality holds throughout the first line only if $\pi_1 = 0$ or $\nu^\ast(P_0, P_1) = 0$, and similarly for the second line.
\end{cor}
\begin{IEEEproof}
The left inequality in the first line follows from the first line of Lemma~\ref{lem:maxMixProp} by adding $\pi_1 (1 - \nu^\ast(P_0, P_1))$ to the denominator, while the second inequality in the first line follows from subtracting $\pi_1 \nu^\ast(P_0, P_1))$ from the denominator. 
\end{IEEEproof}

Given condition \eqref{eqn:contamBound}, the contamination model \eqref{eqn:Ptilde01} has an equivalent representation as specified by \cite[Lem.~1]{scott2013COLT}:
\begin{subequations}\label{eqn:Ptilde01Alt}
\begin{align}
\tP_0 &= (1 - \tpi_0) P_0 + \tpi_0 \tP_1, \qquad \tpi_0 = \frac{\pi_0}{1-\pi_1} \in [0,1),\label{eqn:Ptilde0Alt}\\
\tP_1 &= (1-\tpi_1) P_1 + \tpi_1 \tP_0, \qquad \tpi_1 = \frac{\pi_1}{1-\pi_0} \in [0,1).\label{eqn:Ptilde1Alt}
\end{align}
\end{subequations}
This alternative form makes clear that once $(\tP_0, \tP_1)$ and the modified parameters $(\tpi_0, \tpi_1)$ (or equivalently $(\pi_0, \pi_1)$) are fixed, $(P_0, P_1)$ are also specified exactly. 
Using \eqref{eqn:Ptilde01Alt}, \cite[Cor.~1]{scott2013COLT} shows that $\tpi_0$ and $\tpi_1$ are uniquely determined 
under the irreducibility conditions $\nu^\ast(P_0, \tP_1) = \nu^\ast(P_1, \tP_0) = 0$.  The next lemma provides general expressions for $\tpi_0$, $\tpi_1$ that do not require irreducibility.
\begin{lemma}\label{lem:pitilde01}
The contamination model \eqref{eqn:Ptilde01Alt} has a unique solution in $(\tpi_0, \tpi_1)$ in terms of maximal mixture proportions:
\begin{align*}
\tpi_0 &= \frac{\nu^\ast(\tP_0, \tP_1) - \nu^\ast(P_0, \tP_1)}{1 - \nu^\ast(P_0, \tP_1)},\\
\tpi_1 &= \frac{\nu^\ast(\tP_1, \tP_0) - \nu^\ast(P_1, \tP_0)}{1 - \nu^\ast(P_1, \tP_0)}.
\end{align*}
\end{lemma}
\begin{IEEEproof}
By \cite[Prop.~2]{scott2013COLT} (originally \cite[Prop.~5]{blanchard2010}), there exists a distribution $P_0'$ such that $\nu^\ast(P_0', \tP_1) = 0$ and 
\begin{equation}\label{eqn:P0'1}
\tP_0 = (1 - \nu^\ast(\tP_0, \tP_1)) P_0' + \nu^\ast(\tP_0, \tP_1) \tP_1.
\end{equation}
(An explicit construction for $P_0'$ is given in the proof of \cite[Prop.~5]{blanchard2010}.)  Combining \eqref{eqn:P0'1} with \eqref{eqn:Ptilde0Alt} and solving for $P_0$, we have 
\begin{equation}\label{eqn:P0'2}
P_0 = \frac{1 - \nu^\ast(\tP_0, \tP_1)}{1 - \tpi_0} P_0' + \frac{\nu^\ast(\tP_0, \tP_1) - \tpi_0}{1 - \tpi_0} \tP_1,
\end{equation}
noting that $\tpi_0 < 1$.  From definition \eqref{eqn:maxMixProp} and \eqref{eqn:Ptilde0Alt}, it is seen that both coefficients in \eqref{eqn:P0'2} are non-negative and sum to $1$. Hence \eqref{eqn:P0'2} is a valid mixture decomposition of $P_0$ into $P_0'$ and $\tP_1$.  Furthermore, since $\nu^\ast(P_0', \tP_1) = 0$, we may apply \cite[Cor.~1]{scott2013COLT} to \eqref{eqn:P0'2} to obtain 
\[
\frac{1 - \nu^\ast(\tP_0, \tP_1)}{1 - \tpi_0} = 1 - \nu^\ast(P_0, \tP_1).
\]
Solving for $\tpi_0$ results in the first line in the lemma statement.  The expression for $\tpi_1$ is similarly obtained.
\end{IEEEproof}

Combining Lemmas~\ref{lem:maxMixProp} and \ref{lem:pitilde01} yields a characterization of the contamination proportions $\pi_0$, $\pi_1$. 
\begin{theorem}\label{thm:pi01}
Under condition~\eqref{eqn:contamBound}, we have the relations 
\begin{align*}
\pi_0 + \nu^\ast(\tP_0, \tP_1) \pi_1 &= \frac{\nu^\ast(\tP_0, \tP_1) - \nu^\ast(P_0, P_1)}{1 - \nu^\ast(P_0, P_1)},\\
\nu^\ast(\tP_1, \tP_0) \pi_0 + \pi_1 &= \frac{\nu^\ast(\tP_1, \tP_0) - \nu^\ast(P_1, P_0)}{1 - \nu^\ast(P_1, P_0)}.
\end{align*}
\end{theorem}
\begin{IEEEproof}
We substitute the first line of Lemma~\ref{lem:maxMixProp} into the first line of Lemma~\ref{lem:pitilde01} to obtain 
\[
\tpi_0 = \frac{\bigl(1 - \pi_1 + \pi_1 \nu^\ast(P_0, P_1)\bigr) \nu^\ast(\tP_0, \tP_1) - \nu^\ast(P_0, P_1)}{1 - \pi_1 + \pi_1 \nu^\ast(P_0, P_1) - \nu^\ast(P_0, P_1)}.
\]
Using \eqref{eqn:Ptilde0Alt} and rearranging numerator and denominator,
\begin{align*}
\frac{\pi_0}{1-\pi_1} &= \frac{\nu^\ast(\tP_0, \tP_1) - \nu^\ast(P_0, P_1) - \pi_1 \nu^\ast(\tP_0, \tP_1) (1 - \nu^\ast(P_0, P_1))}{(1 - \pi_1)(1 - \nu^\ast(P_0, P_1))},\\
\pi_0 &= \frac{\nu^\ast(\tP_0, \tP_1) - \nu^\ast(P_0, P_1)}{1 - \nu^\ast(P_0, P_1)} - \pi_1 \nu^\ast(\tP_0, \tP_1),
\end{align*}
which is equivalent to the first relation in the theorem statement. The derivation of the second relation is again analogous.
\end{IEEEproof}

Since $\tP_0$, $\tP_1$ and hence $\nu^\ast(\tP_0, \tP_1)$, $\nu^\ast(\tP_1, \tP_0)$ are assumed to be known, Theorem~\ref{thm:pi01} can be interpreted as a system of equations relating $\pi_0$, $\pi_1$ to the maximal proportions $\nu^\ast(P_0, P_1)$, $\nu^\ast(P_1, P_0)$ for the pure distributions.  If $\nu^\ast(\tP_0, \tP_1)$, $\nu^\ast(\tP_1, \tP_0) < 1$, i.e., if $\tP_0 \neq \tP_1$, then this system is invertible because the determinant $1 - \nu^\ast(\tP_0, \tP_1) \nu^\ast(\tP_1, \tP_0) > 0$, and Theorem~\ref{thm:pi01} describes a bijection.

\begin{figure}
\centering
\includegraphics[width=0.4\linewidth]{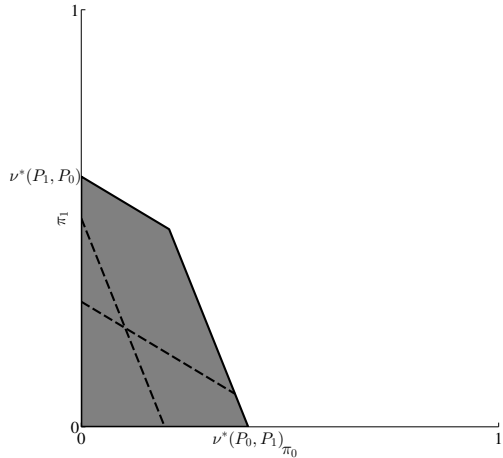}
\caption{Region of feasible contamination proportions $(\pi_0, \pi_1)$ given contaminated distributions $\tP_0$ and $\tP_1$.}
\label{fig:feasiblepi}
\end{figure}

Fig.~\ref{fig:feasiblepi} depicts the set of feasible $(\pi_0, \pi_1)$ values given the contaminated maximal proportions $\nu^\ast(\tP_0, \tP_1)$, $\nu^\ast(\tP_1, \tP_0)$.  The solid outer lines correspond to the mutually irreducible case, namely $\nu^\ast(P_0, P_1) = \nu^\ast(P_1, P_0) = 0$ in Theorem~\ref{thm:pi01}, and the intersection of the lines is the solution characterized in \cite[Prop.~3]{scott2013COLT}.  Theorem~\ref{thm:pi01} generalizes to the interior of the region by specifying solutions for nonzero values of $\nu^\ast(P_0, P_1)$, $\nu^\ast(P_1, P_0)$.  In particular, the dashed lines in Fig.~\ref{fig:feasiblepi} are lines of constant $\nu^\ast(P_0, P_1)$ or $\nu^\ast(P_1, P_0)$ and are parallel to the boundary lines.  This geometry is used in the next section to 
describe uncertainty in $\pi_0$, $\pi_1$.

\section{Contamination-Robust Hypothesis Testing}
\label{sec:robust}

This section discusses the determination of decision rules that are robust to uncertainty in the contamination proportions $\pi_0$ and $\pi_1$.  
Defining $\mbpi = (\pi_0, \pi_1)$, we rewrite the Bayes risk \eqref{eqn:BayesRisk} as follows, 
\begin{equation}\label{eqn:BayesRiskContam}
R_{\Bayes}(\hH,\mbpi) = c_{01} q_0 R_0(\hH,\mbpi) + c_{10} (1-q_0) R_1(\hH,\mbpi),
\end{equation}
to make explicit the dependence on the contamination proportions. 
%
From \eqref{eqn:Ptilde01Alt}, the two error probabilities under the true distributions $P_0$, $P_1$ can be expressed as 
\begin{subequations}\label{eqn:R01}
\begin{align}
R_0(\hH,\mbpi) &= \frac{(1-\pi_1) \tR_0(\hH) - \pi_0(1-\tR_1(\hH))}{1-\pi_0-\pi_1},\label{eqn:R0}\\
R_1(\hH,\mbpi) &= \frac{(1-\pi_0) \tR_1(\hH) - \pi_1(1-\tR_0(\hH))}{1-\pi_0-\pi_1}.\label{eqn:R1}
\end{align}
\end{subequations}
The performance thus depends on the error probabilities $\tR_0(\hH)$, $\tR_1(\hH)$ under the contaminated distributions, which can be determined for fixed decision rule $\hH$, and $\pi_0$, $\pi_1$, which are only partially known.

The set of possible $(\pi_0, \pi_1)$ values is constrained by knowledge of $\tP_0$ and $\tP_1$ as shown in Fig.~\ref{fig:feasiblepi}.  In addition to these initial constraints, we also consider lower and/or upper bounds on $\pi_0$, $\pi_1$ and the maximal mixture proportions $\nu^{\ast}(P_0, P_1)$, $\nu^{\ast}(P_1, P_0)$ for the pure distributions.  As seen from Theorem~\ref{thm:pi01} and Fig.~\ref{fig:feasiblepi}, bounds on $\nu^{\ast}(P_0, P_1)$, $\nu^{\ast}(P_1, P_0)$ correspond to linear inequalities in $\pi_0$, $\pi_1$.  It follows that the feasible region for $(\pi_0, \pi_1)$ is in general a convex polygon, which we may represent as a system of linear inequalities:
\[
\Pi = \{\mbpi : \mba_i^T \mbpi \leq b_i, \; i=1,\dots,m \}
\]
with appropriate choices of $\mba_i \in \mathbb{R}^2$ and $b_i \in \mathbb{R}$. 

The additional bounds on $\pi_0$, $\pi_1$, $\nu^{\ast}(P_0, P_1)$, $\nu^{\ast}(P_1, P_0)$ may be provided by application-specific knowledge and past experience. For example, with voluntary resignation, we can examine the resignation rate historically and use it to roughly characterize or bound $\pi_0$. Moreover, examining data from more than a year in the past, we can observe $P_0$ and $P_1$ without contamination because any employee who was active then and has not resigned yet is by definition not a contaminated sample. Such historical $P_0$ and $P_1$ can be used to bound present values of $\nu^{\ast}(P_0, P_1)$ and $\nu^{\ast}(P_1, P_0)$.\footnote{One may ask why historical $P_0$ and $P_1$ cannot simply be used to determine the decision rule in the present; this is not possible in dynamic business environments where the resignation rate within job roles, skill sets, professions, and organizational units\hspace{1pt}---\hspace{1pt}which are all observations to predict resignation\hspace{1pt}---\hspace{1pt}changes rapidly due to technology trends and management changes. It is the level of differentiation between the classes that we assume does not change much over time, allowing us to bound $\nu^{\ast}(P_0, P_1)$ and $\nu^{\ast}(P_1, P_0)$.} In the case of finding suitable internal candidates for openings, similar openings filled in adjacent groups can provide bounds on $\pi_0$, $\pi_1$, $\nu^{\ast}(P_0, P_1)$, $\nu^{\ast}(P_1, P_0)$.

In this paper, the decision rule $\hH$ is chosen to minimize the Bayes risk subject to worst-case uncertainty in $(\pi_0, \pi_1)$ within the set $\Pi$:
\begin{equation}\label{eqn:BayesTest}
\hH_{\Bayes} = \argmin_{\hH} \, \max_{\mbpi \in \Pi} \, R_{\Bayes}(\hH,\mbpi).
\end{equation}
Alternative formulations include minimizing the worst-case deviation from the true Bayes risk (instead of the absolute Bayes risk in \eqref{eqn:BayesTest}) and minimizing the average Bayes risk over $\Pi$ with respect to some distribution for $\mbpi$.  We leave these alternatives for future work.

The inner maximization in \eqref{eqn:BayesTest} can be restricted to a subset of the vertices of $\Pi$.  For a vertex $\mbpi \in \Pi$, define $I(\mbpi) \subseteq \{1,\dots,m\}$ to be the set of constraints $\mba_i^T \mbpi \leq b_i$ that are met with equality (active constraints), and $\cone\left(\{\mba_i, i \in I(\mbpi)\}\right)$ to be the cone formed by non-negative combinations of the corresponding $\mba_i$. We use $\mathbb{R}_{-}^2$ as a shorthand for the non-positive quadrant of $\mathbb{R}^2$.
\begin{lemma}\label{lem:vertexBayes}
Assume that $\hH$ satisfies $\tR_0(\hH) + \tR_1(\hH) \leq 1$.  Let $\mbpi^{k}$, $k = 1,\dots,V$, be the vertices of $\Pi$ such that 
\begin{equation}\label{eqn:coneCond}
\cone\bigl(\{\mba_i, i\in I(\mbpi^k) \}\bigr) \cap \mathbb{R}_{-}^2 \neq \emptyset.
\end{equation}
Then
\[
\max_{\mbpi \in \Pi} \, R_{\Bayes}(\hH,\mbpi) = \max_{k=1,\dots,V} \, R_{\Bayes}(\hH,\mbpi^k).
\]
\end{lemma}
\begin{IEEEproof}
The restriction to vertices of $\Pi$ follows from the fact that $R_{\Bayes}(\hH,\mbpi)$ is a linear-fractional function of $\mbpi$ for fixed $\hH$. 
This property is seen by substituting \eqref{eqn:R01} into \eqref{eqn:BayesRiskContam} to obtain 
\begin{equation}\label{eqn:BayesRiskQuasilinear}
R_{\Bayes}(\hH,\mbpi) = \frac{\mbc^T \mbpi + d}{1 - \pi_0 - \pi_1},
\end{equation}
where $\mbc \in \mathbb{R}^2$ and $d \in \mathbb{R}$ do not depend on $\mbpi$ (explicit expressions are omitted here).  
Given \eqref{eqn:BayesRiskQuasilinear}, the maximization of $R_{\Bayes}(\hH,\mbpi)$ may be carried out as a search for the largest $t \geq 0$ for which the linear program
\begin{equation}\label{eqn:LPBayes}
\max_{\mbpi \in \Pi} \, \mbc^T \mbpi + d - t(1 - \pi_0 - \pi_1)
\end{equation}
has a non-negative optimal value, implying that the superlevel set $\{\mbpi \in \Pi : R_{\Bayes}(\hH,\mbpi) \geq t \}$ is non-empty.  Since \eqref{eqn:LPBayes} is a linear optimization over a bounded polygon, there exists a vertex of $\Pi$ that is optimal \cite[Thm.~2.8]{bt1997}.  This holds in particular for $t = \max_{\mbpi \in \Pi} R_{\Bayes}(\hH,\mbpi)$ and hence it is sufficient to consider only the vertices of $\Pi$ in maximizing $R_{\Bayes}(\hH,\mbpi)$.

The restriction to vertices satisfying \eqref{eqn:coneCond} is due to the KKT optimality condition for the maximization of $R_{\Bayes}(\hH,\mbpi)$:
\begin{equation}\label{eqn:KKTBayes}
\nabla_{\mbpi} R_{\Bayes}(\hH,\mbpi) = \sum_{i\in I(\mbpi)} \mu_i \mba_i, \quad \mu_i \geq 0,
\end{equation}
which is a necessary condition because $\Pi$ is defined by linear inequalities \cite[Prop.~3.3.7]{bertsekas1999}.  Using \eqref{eqn:R0}, \eqref{eqn:linFracDeriv}, 
and the assumption $\tR_0(\hH) + \tR_1(\hH) \leq 1$, we find that 
\begin{align*}
\frac{\partial R_0(\hH,\mbpi)}{\partial\pi_0} &= -\frac{(1-\pi_1)\bigl(1 - \tR_0(\hH) - \tR_1(\hH)\bigr)}{(1 - \pi_0 - \pi_1)^2} \leq 0,\\
\frac{\partial R_0(\hH,\mbpi)}{\partial\pi_1} &= -\frac{\pi_0 \bigl(1 - \tR_0(\hH) - \tR_1(\hH)\bigr)}{(1 - \pi_0 - \pi_1)^2} \leq 0,
\end{align*}
and similarly for $R_1(\hH,\mbpi)$.  Since $R_{\Bayes}(\hH,\mbpi)$ is a non-negative combination of $R_0(\hH,\mbpi)$ and $R_1(\hH,\mbpi)$ from \eqref{eqn:BayesRiskContam}, we have $\nabla_{\mbpi} R_{\Bayes}(\hH,\mbpi) \in \mathbb{R}_{-}^2$ in \eqref{eqn:KKTBayes}, while the right-hand side of \eqref{eqn:KKTBayes} can range over $\cone\bigl(\{\mba_i, i\in I(\mbpi) \}\bigr)$.  We conclude that it suffices to consider vertices satisfying \eqref{eqn:coneCond}.
\end{IEEEproof}
\begin{remark}
The condition $\tR_0(\hH) + \tR_1(\hH) \leq 1$ is satisfied by any decision rule $\hH$ that is at least as good as random guessing. Hence no generality is lost. 
\end{remark}

Combining \eqref{eqn:BayesTest} and Lemma~\ref{lem:vertexBayes} yields 
\begin{equation}\label{eqn:BayesTestVertex}
\hH_{\Bayes} = \argmin_{\hH} \, t \quad \text{s.t.} \quad R_{\Bayes}(\hH,\mbpi^k) \leq t, \quad k=1,\dots,V.
\end{equation}
In the two-dimensional case considered here, the number $V$ of vertices satisfying \eqref{eqn:coneCond} is very small and $\mbpi^1,\dots,\mbpi^V$ are easily enumerated.  Therefore \eqref{eqn:BayesTestVertex} represents a significant simplification compared to \eqref{eqn:BayesTest}.  However, enumeration becomes increasingly difficult in higher dimensions that would arise in hypothesis testing with more than two hypotheses.

\section{Numerical Examples}
\label{sec:numerical}

In this section we illustrate the proposed minimax procedure via two examples with likelihoods that are not mutually irreducible: Gaussian distributions with different means and different variances, and exponential distributions with different inverse scale parameters.  The Gaussians example provides a rough model for features that predict voluntary resignation, since features such as time since the last job promotion and annual performance rating tend to be approximately normal in many organizations.  The exponentials example provides a rough model for abilities among a high-performing group, which arises when finding suitable candidates.

Consider $P_0 \sim \mathcal{N}(\mu_0,\sigma_0^2)$ and $P_1 \sim \mathcal{N}(\mu_1,\sigma_1^2)$ where $\mu_0 \neq \mu_1$ and, without loss of generality, $\sigma_0 < \sigma_1$.  For this problem, the uncontaminated error probabilities for a likelihood ratio test with threshold value $\gamma$ are:
\begin{align*}
R_0(\gamma) &= Q\left(\tfrac{y^+ - \mu_0}{\sigma_0}\right) + Q\left(\tfrac{-y^- + \mu_0}{\sigma_0}\right) \\
R_1(\gamma) &= 1 - Q\left(\tfrac{y^+ - \mu_1}{\sigma_1}\right) - Q\left(\tfrac{-y^- + \mu_1}{\sigma_1}\right),
\end{align*}
where $Q(y) = \frac{1}{\sqrt{2\pi}}\int_y^\infty \exp(-y'^2/2)dy'$, and $y^+$ and $y^-$ are the solutions to the quadratic equation:
\begin{displaymath}
(\sigma_1^2 - \sigma_0^2)y^2 + 2(\mu_1\sigma_0^2 - \mu_0\sigma_1^2)y + \mu_0^2\sigma_1^2 - \mu_1^2\sigma_0^2 - 2\sigma_0^2\sigma_1^2\ln\left(\gamma\tfrac{\sigma_1}{\sigma_0}\right) = 0.
\end{displaymath}

We examine the situation in which $\mu_0 = 0$, $\mu_1 = 0.2$, $\sigma_0 = 1$, and $\sigma_1 = 2$.  Additionally, for the Bayes risk, we consider the simple case when $q_0 = 0.5$ and $c_{01} = c_{10} = 1$.  The true contamination proportions, unknown to an observer, are $\pi_0 = 0.2$ and $\pi_1 = 0.3$.  These contamination proportions result in $\nu^\ast(\tP_0, \tP_1) = 0.2857$ and $\nu^\ast(\tP_1, \tP_0) = 0.7202$, which are observed.  Additional information on the contamination gives us the constraints $\pi_0 \ge 0.05$ and $\pi_1 \ge 0.1$, as well as $\pi_0 + \nu^\ast(\tP_0, \tP_1)\pi_1 \ge 0.2$ and $\nu^\ast(\tP_1, \tP_0)\pi_0 + \pi_1 \ge 0.25$.  The last two inequalities follow from Theorem~\ref{thm:pi01} and upper bounds on $\nu^\ast(P_0,P_1)$, $\nu^\ast(P_1,P_0)$. With these constraints, the polygon $\Pi$ has six vertices.

\begin{figure}
\centering
\begin{tabular}{c}
\includegraphics[width=0.4\linewidth]{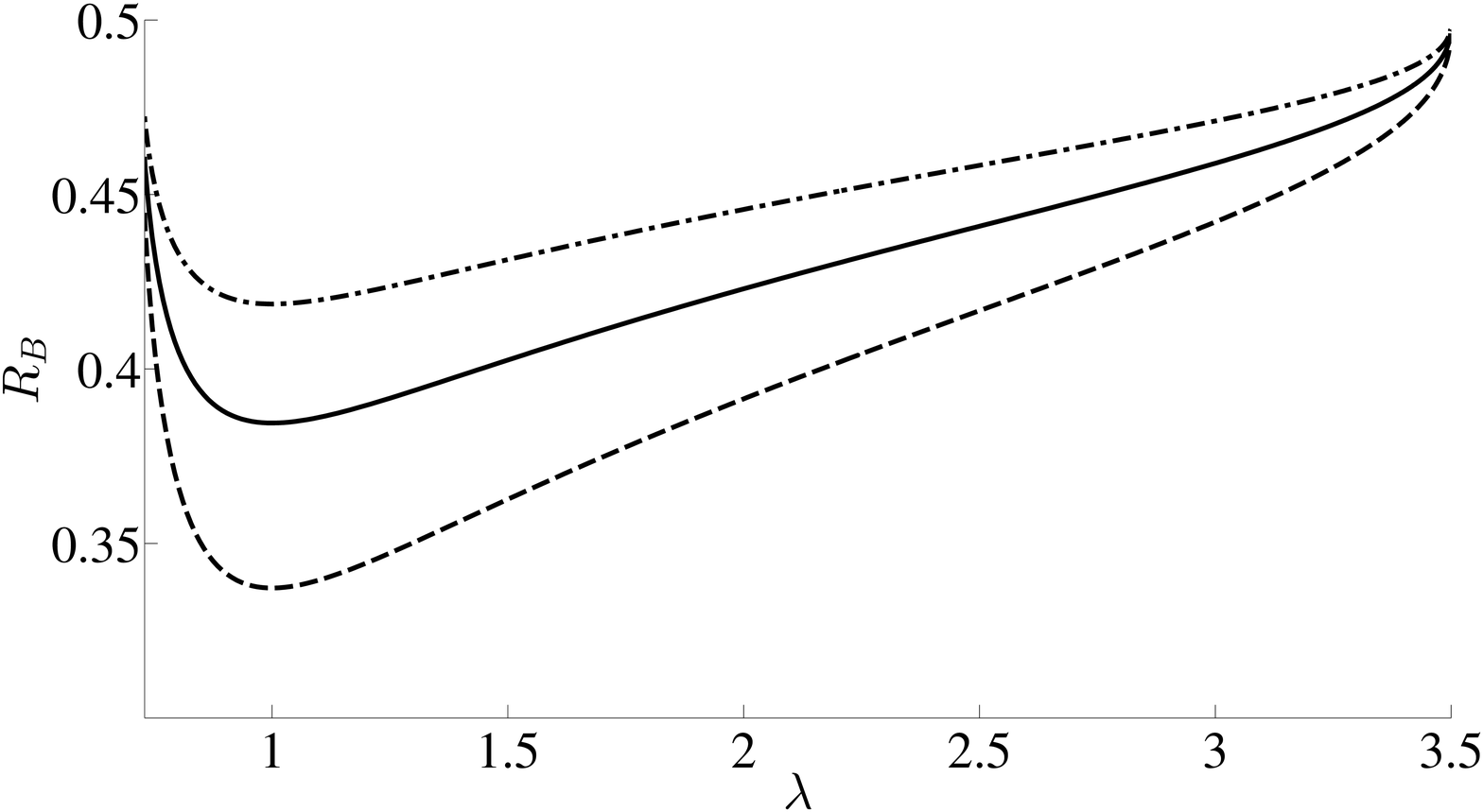}\\
\footnotesize{(a)}\\
\includegraphics[width=0.4\linewidth]{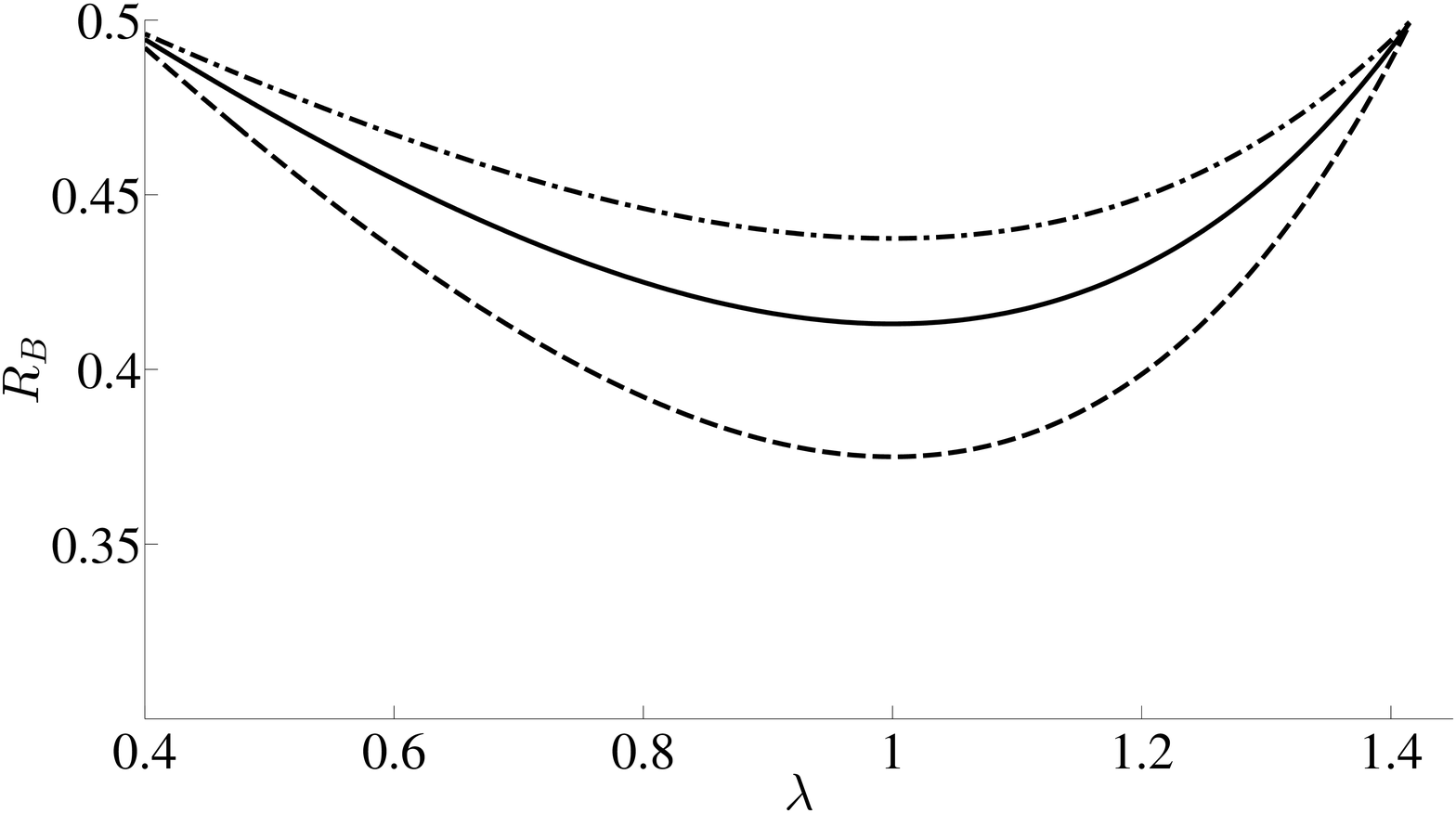}\\
\footnotesize{(b)}
\end{tabular}
\caption{Bayes risk as a function of the threshold on the contaminated likelihood ratio 
for (a) Gaussian example and (b) exponential example: using unknown true contamination proportions (dashed), max solution (solid), and $(0,0)$ contamination proportions (dash-dot).}
\label{fig:RB}
\end{figure}

After performing the inner maximization of the minimax procedure, we find the vertex of $\Pi$ that maximizes the Bayes risk to be $(0.1619, 0.1334)$. 
This maximum Bayes risk is shown in Fig.~\ref{fig:RB}(a) as a function of the threshold $\lambda$ applied to the contaminated likelihood ratio 
($\lambda$ is related to $\gamma$ through a transformation derived in \cite{scott2013COLT}).
The minimum value of this function, i.e.\ the minimax Bayes risk we seek, is $0.3845$.

The figure also shows the Bayes risk if we use the unknown true contamination proportions (which equals the uncontaminated Bayes risk) and the Bayes risk if we use the $(0,0)$ point, i.e., we do not account for contamination.  The minimum Bayes risk using the true contamination proportions is $0.3372$ and the minimum when using $(0,0)$ is $0.4186$.  The minimax solution is between these two values.  Notably, it is less pessimistic than the default $(0,0)$ solution.  The 
solution under irreducibility \cite{scott2013COLT} is not selected under the minimax criterion as it is too optimistic about the Bayes risk value.

As a second example, consider $P_0 \sim \mathcal{E}(\alpha_0)$ and $P_1 \sim \mathcal{E}(\alpha_1)$ where without loss of generality, $\alpha_0 < \alpha_1$.  For this problem, the uncontaminated error probabilities for a likelihood ratio test threshold value $\gamma$ are: $R_0(\gamma) = 1 - e^{-\alpha_0y^*}$ and $R_1(\gamma) = e^{-\alpha_1y^*}$, where 
$y^* = \ln\left(\frac{\alpha_0}{\alpha_1}\gamma\right)/(\alpha_0 - \alpha_1)$.  We set $\alpha_0 = 1$ and $\alpha_1 = 2$ and keep all other parameters the same as in the first example.  With these exponential likelihoods and parameter settings, $\nu^\ast(\tP_0, \tP_1) = 0.7059$ and $\nu^\ast(\tP_1, \tP_0) = 0.3750$ and the resulting $\Pi$ has five vertices.  The maximizing vertex is $(0.1619,0.1334)$ and the maximum Bayes risk is shown in Fig.~\ref{fig:RB}.  The minimax Bayes risk is $0.4130$, which lies between the minimum Bayes risk with known contamination proportions, $0.3750$, and the minimum Bayes risk using proportions $(0,0)$, $0.4375$, in the same manner as the previous example.

\section{Conclusion}
\label{sec:conclusion}

In this paper, we have examined the problem of contaminated likelihood functions that arise due to 
label noise in hypothesis testing.  In contrast to previous work on the subject which derived consistency results for the case when the likelihoods are mutually irreducible, we deal with arbitrary likelihoods and obtain 
decision rules robust to uncertainty in the contamination proportions.  Toward this end, we have posed an optimization problem that is naturally subject to linear constraints and shown that its objective function is a linear-fractional function.  Therefore, the optimization problem reduces to 
linear programs that can be 
simplified using the KKT conditions into 
a search over certain vertices of the constraint set. 
We have shown the method on two numerical examples.

\bibliographystyle{IEEEbib}
\bibliography{contamICASSP}

\end{document}